\documentclass[runningheads]{llncs}
\usepackage[T1]{fontenc}
\usepackage{graphicx}
\usepackage{subcaption}
\usepackage{amsmath}

\usepackage{hyperref} 
\hypersetup{
colorlinks   = true,
citecolor    = blue,
urlcolor    =  blue
}
\newcommand{\footURL}[1]{\footnote{\url{#1}}}

\usepackage[labelformat=simple]{subcaption}

\begin{document}
%
\title{Evolution of Cooperation in LLM-Agent Societies: A Preliminary Study Using Different Punishment Strategies}
\titlerunning{Evolution of Cooperation in LLM-Agent Societies}
%

\author{Kavindu Warnakulasuriya\inst{1}\thanks{Co-first authors.}\orcidID{0009-0002-2462-147X} \and
Prabhash Dissanayake\inst{1}\textsuperscript{*}\orcidID{0009-0002-9192-3013} \and
Navindu De Silva\inst{1}\textsuperscript{*}\orcidID{0009-0009-9190-4753} \and 
Stephen Cranefield\inst{2}\orcidID{0000-0001-5638-1648} \and
Bastin Tony Roy Savarimuthu\inst{2}\orcidID{0000-0003-3213-6319} \and
Surangika Ranathunga\inst{3}\orcidID{0000-0003-0701-0204} \and
Nisansa de Silva\inst{1}\orcidID{0000-0002-5361-4810}
}
\authorrunning{Warnakulasuriya et al.}
%
\institute{University of Moratuwa, Moratuwa, Sri Lanka\\
\email{\{kavinduw.20, prabhash.20, navindu.20, NisansaDdS\}@cse.mrt.ac.lk}
\and
University of Otago, Dunedin, New Zealand\\
\email{\{stephen.cranefield, tony.savarimuthu\}@otago.ac.nz}
\and
Massey University, Auckland, New Zealand\\
\email{S.Ranathunga@massey.ac.nz}}
\maketitle              
\begin{abstract}
The evolution of cooperation has been extensively studied using abstract mathematical models and simulations. Recent advances in Large Language Models (LLMs) and the rise of LLM agents have demonstrated their ability to perform social reasoning, thus providing an opportunity to test the emergence of norms in more realistic agent-based simulations with human-like reasoning using natural language. In this research, we investigate whether the cooperation dynamics presented in Boyd and Richerson's model persist in a more realistic simulation of the Diner's Dilemma using LLM agents compared to the abstract mathematical nature in the work of Boyd and Richerson. Our findings indicate that agents follow the strategies defined in the Boyd and Richerson model, and explicit punishment mechanisms drive norm emergence, reinforcing cooperative behaviour even when the agent strategy configuration varies. Our results suggest that LLM-based Multi-Agent System simulations, in fact, can replicate the evolution of cooperation predicted by the traditional mathematical models. Moreover, our simulations extend beyond the mathematical models by integrating natural language-driven reasoning and a pairwise imitation method for strategy adoption, making them a more realistic testbed for cooperative behaviour in MASs.

\keywords{Multi-Agent Systems \and Large Language Model \and LLM Agent \and Social Dilemmas \and Agent Strategies}
\end{abstract}
\section{Introduction}

Autonomous agents have gained significant popularity due to their immense utility in various real-world applications in customer service, healthcare, social networks, and retail domains~\cite{si_cooperative_2024}. Autonomous agents can even co-exist with humans in virtual environments~\cite{ranathunga2011interfacing}. Multi-agent systems (MASs) bring together agents with independent objectives and decision-making abilities that interact within a shared environment. As such, agents must cooperate and coordinate their actions in dynamic and often unpredictable settings~\cite{savarimuthu2024harnessing}. Cooperative agents can help improve the performance of individual agents and the overall system~\cite{kraus1997negotiation}. Researchers have sought to understand how cooperation emerges in societies and have presented various mathematical models and simulations that predict agent behaviours~\cite{axelrod1986evolutionary,boyd_punishment_1992}. However, the suitability of these models for real-world, human-oriented environments remains uncertain. With the rise of generative AI technologies, traditional rule-based decision-making approaches are being challenged, necessitating further investigation~\cite{mumuni_large_2025}. 

Mathematical approaches such as game-theory models and evolutionary dynamics are commonplace approaches to modelling and predicting agent behaviour \cite{mao_alympics_2024,fontana_nicer_2024}. Often, these models rely on simplified abstractions, particularly in social dilemmas such as the Prisoner's Dilemma~\cite{kollock_social_1998} and the $n$-player Diner's Dilemma~\cite{teng2013trust}, which focus on agent cooperation. On the other hand, social norms are crucial in guiding agents towards cooperative standards~\cite{ren_emergence_2024}. Ensuring adherence to these norms requires punishment to serve as reinforcement for cooperative agents and to penalise defectors~\cite{axelrod1986evolutionary,wang2023emergence}. 

Boyd and Richerson's (B\&R) model~\cite{boyd_punishment_1992}, a prominent simulation study of the evolution of cooperation, suggests that punishment-based mechanisms can sustain long-term cooperation. While being a simple abstract mathematical simulation, its applicability in a more realistic human-based environment remains an underexplored area. With the opportunity of using Large Language Model (LLM) agents, a form of generative AI technology, which demonstrate a great understanding of natural language~\cite{savarimuthu2024harnessing}, we explore different strategy compositions introduced by the B\&R model to examine whether a more realistic simulation of the evolution of cooperation using LLM agents produces similar norm emergence as the abstract B\&R model~\cite{boyd_punishment_1992}.

We model a realistic $n$-player Diner's Dilemma, using LLM agents as the backbone in making the complex dilemma decisions and allowing them to act based on their strategies, calculate payoffs for their dilemma actions, and finally reflect on their actions, analyze other agents, and change their strategies by comparing their utilities. Furthermore, following the B\&R model, we allow the agent population to have multiple strategies in each simulation (currently up to four strategies per simulation). The B\&R model includes experiments with both two and three strategies, but their mathematical model of population dynamics is challenging to extend beyond a small number of strategies. Furthermore, the meaning of the B\&R strategies is directly encoded within their equations. In contrast, our LLM-based approach allows reasoning about strategies using natural language, making it easier to introduce and test alternative strategies in future studies. Therefore, in this study, we choose to allow different combinations of four strategies in the population, making the simulation more complex and realistic, enabling a deeper analysis of the strategy evolution. 

In summary, we aim to investigate the impact of strategies from B\&R's work and their evolution with repeated Diner's Dilemma scenario, modelled with novel LLM agents, which have been shown to implicitly capture human reasoning and thinking abilities, thus allowing us to gain insights into whether these LLM agents behave similarly as shown in the abstract mathematical simulation studies. Our findings suggest that LLM agents in a Diner's Dilemma simulation show promising convergence toward cooperative strategies under explicit increasing punishment costs.

The structure of the remainder of this paper is outlined as follows: Section \ref{sec:relWorks} reviews the related works. Section \ref{sec:methodology} elaborates on our proposed methodology approach to model the implementation for the emergence of cooperative agent behavioural strategies. Sections \ref{sec:experimentation} and \ref{sec:results} highlight the preliminary experiments conducted and the results obtained, with a discussion. Finally, Section \ref{sec:conclusion} concludes the paper, along with future directions for research.

\section{Related Works}
\label{sec:relWorks}

\subsection{Game Theory and Social Dilemmas}

Over the years, agent behaviour has been studied through mathematical analysis of dynamics or computational simulations of evolutionary dynamics~\cite{axelrod1986evolutionary,boyd_punishment_1992,mahmoud_establishing_2015,wang2023emergence}. Such models have limitations because they do not explicitly map to a real-world task or scenario and generalize agent interactions in a fixed structure. In other words, they are limited to simple abstractions. These studies use game theory as a framework to model human decision-making in a highly abstract form. 

Social dilemmas are an agent-related strategic concept that motivates the study of the normative behaviour of agents in MASs. A social dilemma occurs when an agent is forced to choose between actions that maximize their personal gain at the expense of the group's collective benefit or actions that promote the collective good but lower their personal benefits~\cite{kollock_social_1998}. This scenario is essentially a conflict between personal and social optimality. As a result, agents who aim to satisfy their short-term self-interests are often characterized as non-cooperative, as they are less likely to choose actions that serve the long-term benefit of the group. These social dilemmas can be explained through game theory, which analyzes how rational agents make decisions when faced with interactions with individuals in competitive or cooperative game environments. Game theory helps identify vital insights in explaining the behaviour of agents under economic, political and social interaction~\cite{Camerer2010gametheory,bonau2017behavegametheory}.

A dilemma requires the agent or individual to decide whether to \textit{cooperate} with or to \textit{defect} against its opponent. Based on these decisions, four main payoff values are defined, as elaborated by Macy and Flache~\cite{macy2002learningdynamics}: \textit{(i) reward (R)}, given when both agents choose to cooperate, \textit{(ii) punishment (P)}, incurred when both agents defect, \textit{(iii) temptation (T)}, where agent defects and unfairly benefits while the opponent cooperates, and \textit{(iv) sucker cost (S)}, where a cooperating agent suffers a loss when the opponent defects. These payoff values form the foundation of various social dilemmas studied in game theory, such as the Prisoner's Dilemma~\cite{noauthor_prisoners_nodate,howley2009nplayerprisoners}, the chicken game~\cite{macy2002learningdynamics}, stag hunt~\cite{si_cooperative_2024} and the trust game~\cite{liebrand1983classification}. These two-player social dilemmas can be generalised into their $n$-player form~\cite{howley2009nplayerprisoners,liebrand1983classification}. In the Diner's Dilemma~\cite{teng2013trust}, a group of agents agrees to split the cost of their meals. However, individual agents may exploit this arrangement by ordering expensive items and transferring a portion of their costs onto the group.

Social dilemmas create a precarious position for norm emergence in multi-agent societies as agents would look to increase their utility through defection and benefiting from the cooperation of others~\cite{boyd_punishment_1992,axelrod1986evolutionary}. Furthermore, agents would be less likely to cooperate towards a collective goal when it is known that other agents would contribute, such as in the public goods game studied in~\cite{wang2023emergence}. This is known as the \textit{free-rider problem}~\cite{sweeney1973experimental}. Hence, a suitable mechanism is necessary to ensure agents do not exploit cooperative behaviours, thereby discouraging cooperation.

\subsection{Metanorms}

With the risk of social dilemmas causing agents to be self-centred without regret or guilt, a higher-order mechanism needs to be in place. \textbf{Metanorms}, first coined by Axelrod~\cite{axelrod1986evolutionary}, are second-order norms that guide agents in responding to norm violations to enforce them among defectors. These will enforce penalties on non-cooperative agents. Thereby, they aim to eliminate the free-rider problem in social dilemmas. There are two main approaches for implementing metanorms in agent simulations, namely, punishment-based~\cite{axelrod1986evolutionary,boyd_punishment_1992,mahmoud_establishing_2015} and indirect reciprocity~\cite{ohtsuki2006leadingeight,okada2020twoovercome,quan2022keeping}.

Punishment-based implementations~\cite{axelrod1986evolutionary} describe how to enhance norm compliance by punishing norm violators and those who fail to punish violators, treating non-punishment as a defection against the multi-agent community. In norm-based models, the establishment of norms relies on agents willing to enforce compliance, as insufficient enforcement and free-riding problems can lead to norm collapse~\cite{mahmoud_establishing_2015}. Here, punishment incurs a cost to the punisher and a larger cost to the punished agent, ensuring that punishments are not applied discriminately. Works by Axelrod~\cite{axelrod1986evolutionary} and Boyd and Richerson~\cite{boyd_punishment_1992} have identified that the population must maintain a population of \textit{punisher} agents to prevent norm violators from overtaking cooperative populations. 

Indirect reciprocity is an alternate approach for implementing metanorms without compelling agents to punish and reduce their short-term utility. Generally, indirect reciprocity relies on reputation scores and information-sharing mechanisms like public and private reputation framework~\cite{quan2022keeping} or gossip~\cite{nowak2005evolution}. In this initial study, we focus on the punishment-based approach outlined in the B\&R model, which serves as the foundation for our research.

\subsection{LLM Agents}

Despite extensive research on the effects of metanorms and punishment mechanisms to induce cooperation, limited work has been conducted in conjunction with LLMs. In the few works applying LLMs to reasoning about social dilemmas, LLMs struggle with such interactions---GPT-4\footURL{https://openai.com/index/gpt-4} has been shown to select actions that maximize its personal gain and fails to coordinate with fellow agents in games such as the Battle of the Sexes~\cite{akata2023playing} and frequently selects uncooperative actions that harshly penalize minor mistakes by opponents. Therefore, although LLMs exhibit strong alignment with human behaviour, they struggle to achieve the high levels of cooperation seen in real-world human interactions. This limitation suggests that LLMs should be carefully evaluated when integrated into social experiments~\cite{fan2024llmrational}.

Recently,  Fontana et al.~\cite{fontana_nicer_2024} provided insights into the capabilities of LLM agents in handling iterated Prisoner's Dilemma games. Using the Llama-2-70B-chat model\footURL{https://huggingface.co/meta-llama/Llama-2-70b-chat-hf}, it was reported that while the LLM did not display defection initially, it required more iterations to achieve a cooperative majority---demonstrat\-ing slower convergence towards cooperation. They found how the defection rate of an agent's opponent also impacts its behaviour.

Representing social dilemmas for LLM agents requires thorough analysis. Traditional multi-agent models use mathematical frameworks such as payoff matrices and cost-benefit values to predict agent behaviour as seen in~\cite{axelrod1986evolutionary,boyd_punishment_1992}. However, the weak arithmetic capabilities of LLMs may affect LLM agents' effective comprehension and processing of such payoff constraints~\cite{wang2024textsim}. One approach was to provide the LLM with the payoffs for each two-player scenario through the prompts as sentences~\cite{fontana_nicer_2024,akata2023playing}. However, with no metanorm and norm implementations within these simulations, LLM agents often do not engage in cooperative strategies due to low repercussions for exhibiting defection~\cite{mahmoud_establishing_2015}.

Normative multi-agent system researchers have started to investigate the capability of LLMs in norm discovery, reasoning and conformance~\cite{savarimuthu2024harnessing}. The work of He et al.~\cite{he_norm_2024} investigated the ability of three LLMs (Llama 2 7B\footURL{https://huggingface.co/meta-llama/Llama-2-7b}, Mixtral 7B\footURL{https://huggingface.co/mistralai/Mixtral-8x7B-v0.1} and ChatGPT-4) to identify norm violations and reported their promise. The work of Haque and Singh~\cite{haqueExtractingNormsContracts2024} demonstrates the promise of ChatGPT in extracting norms from contracts without requiring training or fine-tuning of datasets. However, none of the prior works have investigated the capability of LLMs to promote agents to adopt and imitate the cooperative behavioural strategies seen by other agents. 

\subsection{Agent Simulations}\label{subsect:Agent-Simulations}

Simulations are commonly utilised to explore AI agent behaviour within a virtual environment \cite{wang2024survey}. These frameworks use research from multiple fields, such as social sciences, psychology, economics and AI for understanding social phenomena. To study social behaviour in group settings, it is useful to develop simulations that closely depict human activities and track changes in the world states, such as movements of objects resulting from agent actions~\cite{wang2024textsim}. These simulations are broadly categorized as \textit{task-based} or \textit{social interaction-based} simulations. Task-based simulations~\cite{gu2024groupchat}, such as the \textbf{ScienceWorld} environment, are used for conducting science experiments using a text-based framework~\cite{wang2022scienceworld,ichida2024bdi}, whereas social simulations, such as \textbf{Melting Pot}~\cite{agapiou2022melting}, use multi-agent reinforcement learning environments. The use of LLM agents in simulations provides a more realistic approach in replicating normative behaviours~\cite{leng2023llmsocial,savarimuthu2024harnessing}, as seen in frameworks like \textbf{AgentVerse}~\cite{chen2023agentverse}, but with representation of the objectives in an abstract manner rather than using specific social scenarios. Game engines such as \textbf{ALFWorld}~\cite{shridhar2020alfworld} and \textbf{Watch-And-Help}~\cite{puig2020watchandhelp} provide detailed environmental control but are often complex to modify and lack seamless LLM agent integration. Meanwhile, a sandbox environment, \textbf{Smallville}\footURL{https://github.com/nickm980/smallville}, provides an interactive and customizable LLM agent-driven simulation, which has been utilized to build agent societies~\cite{park2023generative} and also normative frameworks~\cite{ren_emergence_2024}. Thus, we consider that this provides a promising simulation framework for experimenting with the work by B\&R in a realistic manner.

The B\&R model has identified that for cooperation in an $n$-player system to be maintained, there should be a sufficient number of ``Moralist agents'' in the system \cite{boyd_punishment_1992}. However, again, this is simulated in an abstract mathematical way, which is a limitation we intend to address by simulating with LLM agents with a realistic Diner's Dilemma scenario.

In summary, to the best of our knowledge, no research has studied the evolution of cooperation with the strategies introduced in the B\&R work using LLM agents. Although some articles have explored dilemmas, such as the Prisoner's Dilemma with LLMs as well as other strategies, there is a lack of research on exploring the evolutionary aspect of the cooperation strategies within a more realistic, $n$-player dilemma scenario to infer insights on whether it produces similar norm emergence as in the abstract B\&R work.

\section{Methodology}
\label{sec:methodology}

This section outlines the methodology employed to investigate the evolution of cooperation with the Diner's Dilemma scenario implemented through LLM agents. We used the strategy descriptions from the B\&R model, leveraging their provided English descriptions alongside the mathematical formulations to guide the LLM's behaviour, as described in the following section.

\subsection{Simulation Environment Setup}

\begin{figure}[!htb]
    \centering
    \includegraphics[width=1\textwidth]{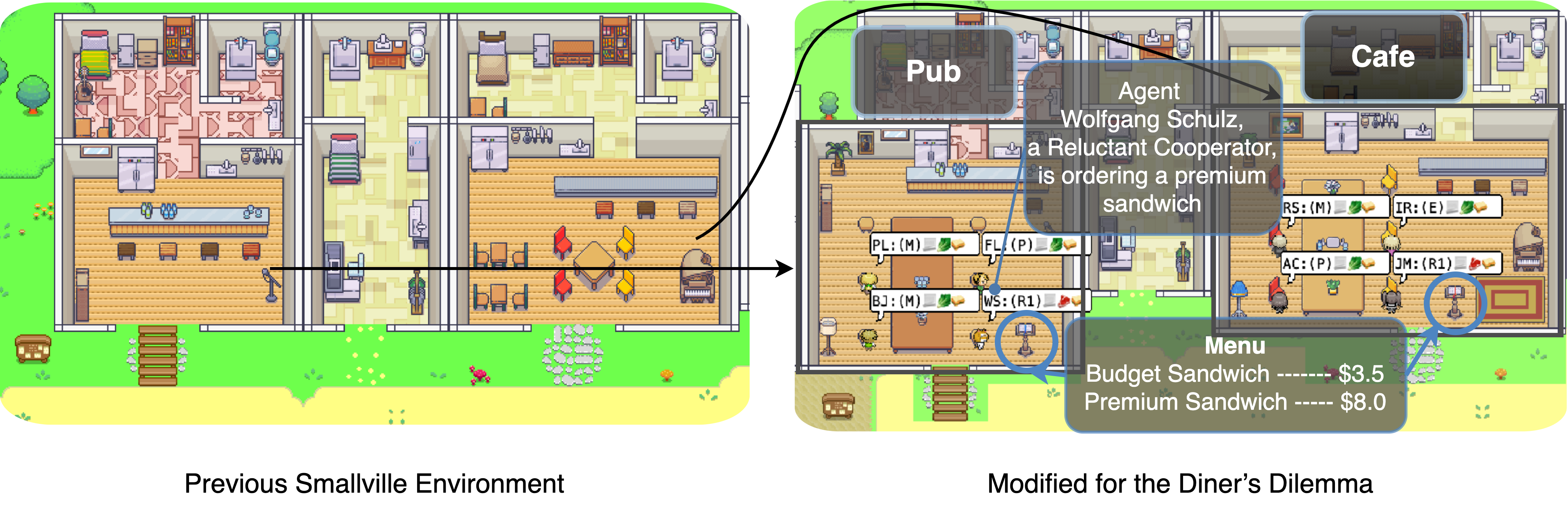}
    \caption{Smallville environment modifications for the Diner's Dilemma simulation}
    \label{fig:smallville_environment_modifications_for_diner's_dilemma_simulation}
\end{figure}

The simulation framework was adapted from the existing Smallville environment utilized in the CRSEC framework~\cite{ren_emergence_2024}\footURL{https://github.com/sxswz213/CRSEC} as it is a promising framework for experimenting with the work of B\&R in a realistic manner, as explained in Section~\ref{subsect:Agent-Simulations}. The framework was utilized to model realistic social dilemmas, specifically, The Diner's Dilemma, as illustrated in Figure~\ref{fig:smallville_environment_modifications_for_diner's_dilemma_simulation}. The virtual environment was designed using the Tiled Map Editor\footURL{https://www.mapeditor.org/} for layout and Phasor\footURL{https://phaser.io/} for agent movement, creating two primary settings: a pub and a cafe, where the agents interact. A total of eight agents were introduced and divided into two groups, where each agent was assigned distinct strategies and lifestyles to simulate diverse attributes and interactions. Agent lifestyles---a concept used in the works of Ren et al.~\cite{ren_emergence_2024}---can be used to simulate human-like behaviour (\textit{e.g., ``Likes to take a high-intensity run in the morning and needs high nutrition for it''}) in our LLM agents and to test for biases in LLM decision prompting when engaging in social dilemmas.

Agents were assigned the following strategies based on Boyd and Richerson's model.
\begin{itemize}
    \item \textbf{Cooperator-Punisher (P):} Always cooperates and punishes defectors.
    \item \textbf{Reluctant Cooperator (R1):} Defects until punished, then cooperates indefinitely (without punishing others).
    \item \textbf{Easy Going Cooperator (E):} Always cooperates but never punishes.
    \item \textbf{Moralist (M):} Always cooperates and punishes defectors and non-punish\-ers, and those who fail to punish non-punishers.
\end{itemize}

In our research, we encode these strategies as part of the prompts provided to the LLM, guiding the agent's decision-making process in social dilemma situations~\cite{noauthor_prisoners_nodate}. As agents navigate the scenario in the simulated environment, we consider a norm to have emerged when a significant proportion of the population adopts a successful strategy transmitted through evolutionary mechanisms. Hence, our objective is to investigate the dynamics that emerge from the interactions of these strategies within the population of the simulated environment.


\subsection{Diner's Dilemma Simulation Process}
\label{subsec3_2:diners_dilemma_sim_pcs}

\begin{figure}[!htb]
    \centering
    \includegraphics[width=0.7\textwidth]{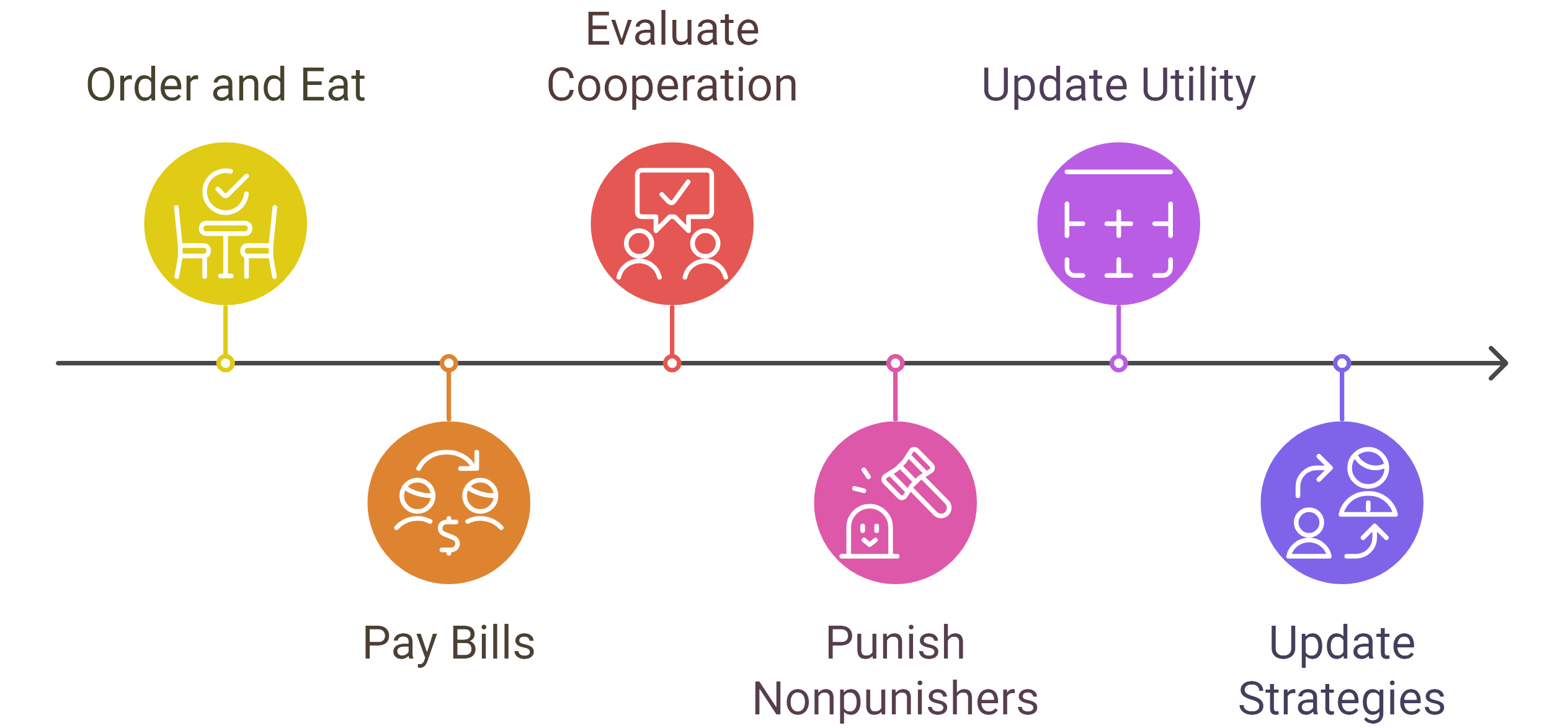}
    \caption{Agent interaction sequence for the Diner's Dilemma. The sequence consists of 6 main processes as shown in the Figure.}
    \label{fig:agent_interaction_sequence}
\end{figure}

The simulation of a Diner's Dilemma involved multiple stages, as illustrated in Figure~\ref{fig:agent_interaction_sequence}, each incorporating LLM-based decision-making processes. In this scenario, a group of agents meet in the cafe or the pub, having agreed to split the cost of their meals. Each agent faces a dilemma in deciding the type of meal to order from the given two options: budget or premium. The agent can either cooperate by choosing the less preferred budget option to increase the collective benefit or choosing the preferred premium (and more expensive) option to maximize the personal gain. Then, the agents apply their individual strategies to determine whether to punish defectors (defined as agents who choose the premium meal) and, in the case of Moralists, to punish those who fail to punish defectors, thereby enforcing metanorms. Next, the agents update their individual utilities according to their decisions in ordering and considering the punishment costs. Finally, the agents compare their utilities with other agents and determine whether to adopt a different strategy. The individual stages are described in detail below.

\begin{enumerate}
    \item \textbf{Dilemma Decision Stage (Order):} Agents are prompted to make decisions regarding their meal orders mainly based on their strategies and the menu provided (budget vs. premium options, where the premium option is more expensive). Other inputs include the names and number of other agents in the group (Figure~\ref{fig:order_eat_and_pay_bills}).
    \begin{figure}[!htb]
        \centering
        \includegraphics[width=1\textwidth]{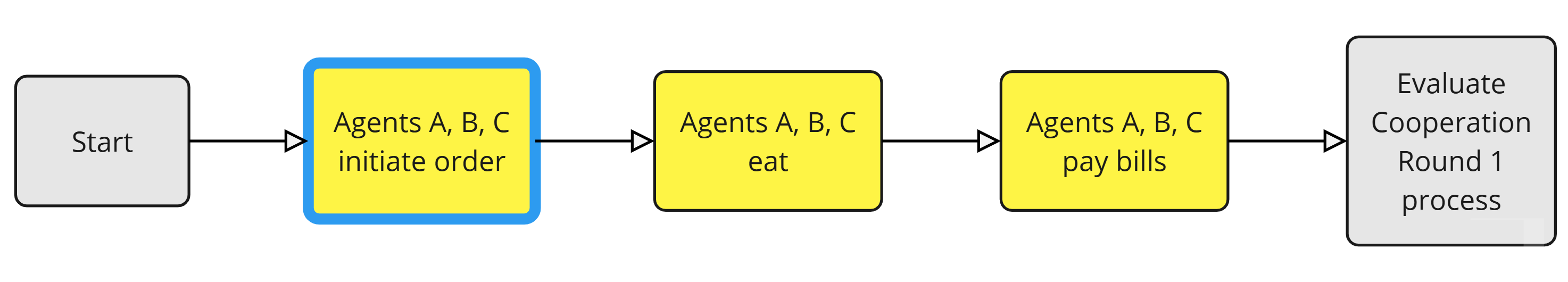}
        \caption{Agent interaction sequence within the Dilemma Decision Stage consisting of the Order \& Eat and Pay Bills processes. The thick blue-outlined block represents the process that prompts the LLM to make a decision.}
        \label{fig:order_eat_and_pay_bills}
    \end{figure}
    
    \item \textbf{Punishment Stage (Evaluate Cooperation - Round 1):} Agents evaluate the actions of others and decide whether to punish defectors. LLMs are used to decide whether to scold one another based on the agent's strategy. Other inputs consist of the order history (Figure~\ref{fig:punishment_for_defectors}).
    \begin{figure}[!htb]
        \centering
        \includegraphics[width=1\textwidth]{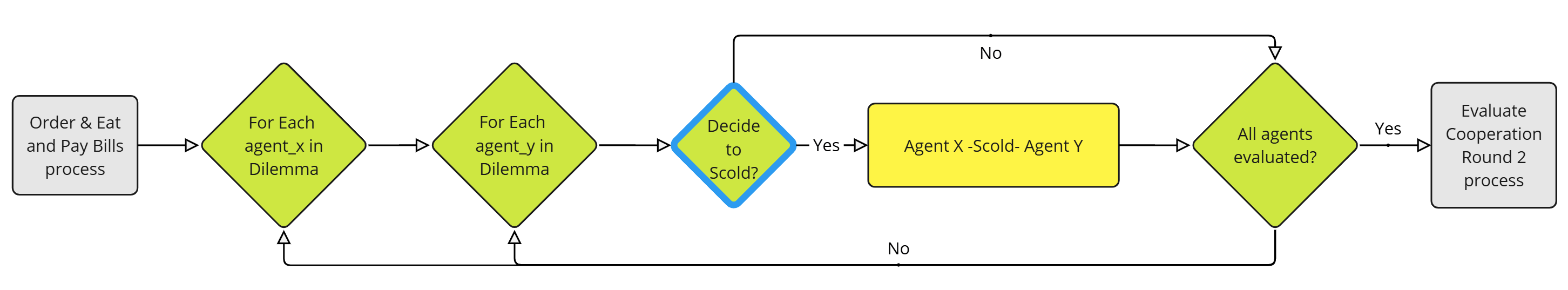}
        \caption{Agent interaction sequence for the Punishment Stage to evaluate cooperation. The thick blue-outlined block represents the process that prompts the LLM to make a decision.}
        \label{fig:punishment_for_defectors}
    \end{figure}
    
    \item \textbf{Metanorm Enforcement Stage (Evaluate Cooperation - Round 2):} This stage involves higher-order normative reasoning (i.e., a metanorm), where moralist agents punish those who have failed to enforce norms (Figure~\ref{fig:punishment_for_non-punishers_and_non-punisher_non-_punisher}).
    \begin{figure}[!htb]
        \centering
        \includegraphics[width=1\textwidth]{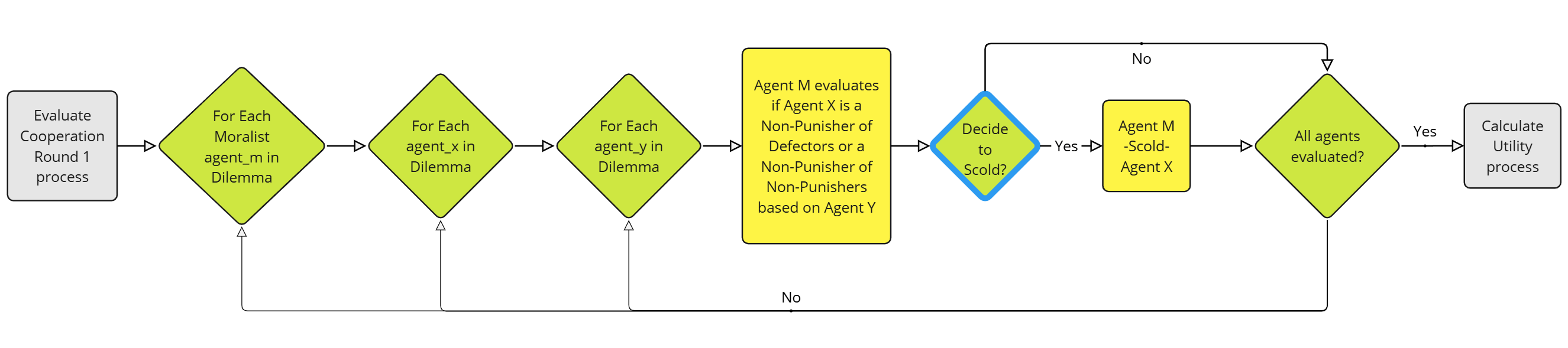}
        \caption{Agent interaction sequence for punishment for the Metanorm Enforcement Stage for the punishment of non-punishers and agents who do not punish non-punishers. The thick blue-outlined block represents the process that prompts the LLM to make a decision.}
        \label{fig:punishment_for_non-punishers_and_non-punisher_non-_punisher}
    \end{figure}
    
    \item \textbf{Utility Assessment and Strategy Update:} Each agent's actions and outcomes are logged to update a numerical utility score that reflects its current performance. Here, apart from the order history, punishment costs are also considered when updating the utility. Punishing another agent will incur a cost of $k$ to the punisher and a cost of $p$ to the agent who is being punished (where usually $p$ > $k$). In small populations, changes in population dynamics can be observed by varying \textit{p} while keeping \textit{k} fixed \cite{boyd_punishment_1992,noauthor_evolution_nodate}. Accordingly, when assigning punishment cost values to the LLM agents, we set \textit{k = 1} and increased the value of \textit{p}. To determine whether an agent should adopt a new strategy, we utilize a pairwise imitation method based on the Fermi function \cite{traulsen_pairwise_2007} to drive the spread of successful strategies. Here, with the payoffs calculated during the Diner's Dilemma, the utility of the focal agent (A) is compared against that of a randomly chosen role model (B) using the following equation from the Fermi process, which computes the probability of agent A changing to use the target agent's strategy.
    \begin{align}
        p=\frac{1}{1+e^{-\beta{(\pi^{i}_{B}-\pi^{i}_{A})}}}
    \end{align}
    Here, $\beta$ acts as the selection temperature parameter---higher values make agents highly sensitive to even minor differences in utility (thus rapidly adopting more successful strategies), while lower values lead to a more gradual response (We used $\beta$ = 1 in our simulations). The parameters $\pi^{i}_{A}$ and $\pi^{i}_{B}$ represent the utility (payoff) values of agents A and B, respectively, based on their accumulated rewards and penalties. This mechanism enables an adaptive evolution of strategies, as agents probabilistically switch to strategies that yield higher payoffs, thereby driving the evolution of cooperation over time (Figure~\ref{fig:update_utility} and Figure~\ref{fig:update_strategies}).
    \begin{figure}[!htb]
        \centering
        \includegraphics[width=0.8\textwidth]{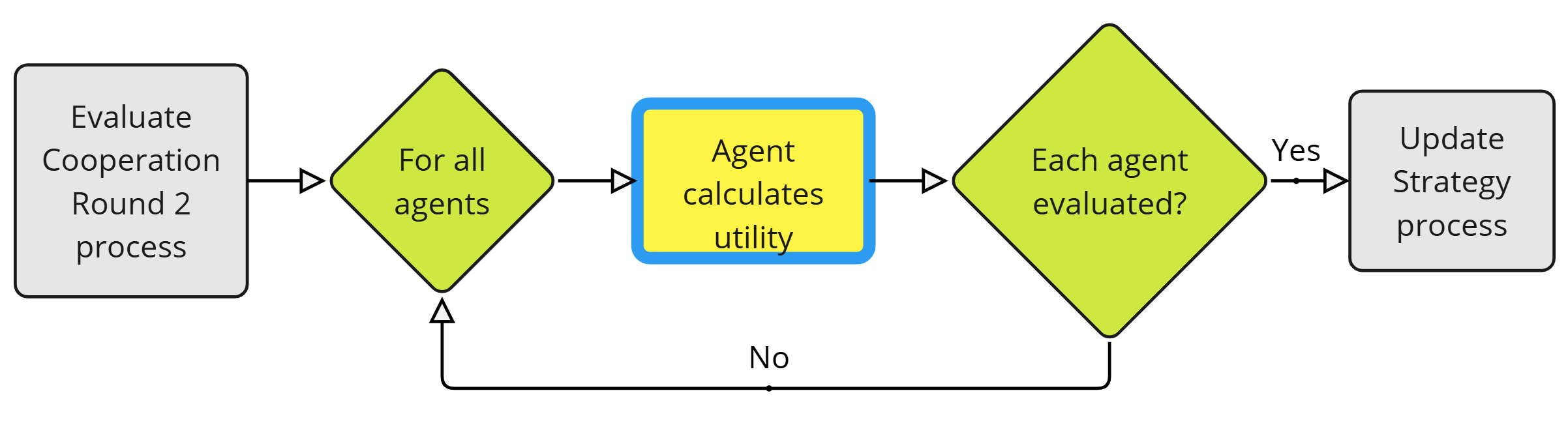}
        \caption{Agent interaction sequence for Utility Assessment. The thick blue-outlined block represents the process that prompts the LLM to make a decision.}
        \label{fig:update_utility}
    \end{figure}
    \begin{figure}[!htb]
        \centering
        \includegraphics[width=1\textwidth]{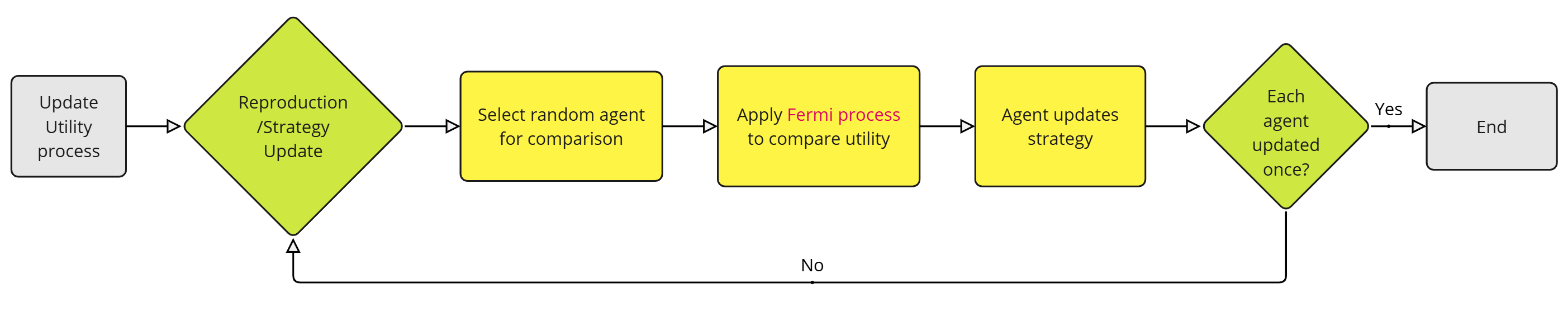}
        \caption{Agent interaction sequence for Strategy Update Process.}
        \label{fig:update_strategies}
    \end{figure}
\end{enumerate}

\section{Experimentation}\label{sec:experimentation}
Our experiments were conducted using the open-source model \textit{Meta-Llama-3.3-70b-Instruct} provided through the Vertex API by Google Cloud without any fine-tuning; we used the original model as provided. We conducted separate testing for each of the four stages where the LLM is used by varying the temperature (randomness of text that is generated by the LLM) and top\_p (or \textit{nucleus sampling}, where tokens with top\_p probability mass are considered), different strategies, and lifestyles of personas (e.g., Morning Runner, Newspaper Reader, Photographer) to assess biases and to ensure expected results are achieved. Finally, we integrated all of the tested prompts and LLM agents with strategies and lifestyles into the simulation.

We conducted 6 total simulations, each with 10 iterations of the Diner's Dilemma scenario. In each simulation, a society of 8 agents was divided into two groups of four, with each group initially meeting at separate locations (the pub and the cafe). After each iteration, the groups will swap the locations (from pub to cafe and vice versa) and engage in the Diner's Dilemma process described in Section~\ref{subsec3_2:diners_dilemma_sim_pcs} iteratively.

Two initial combinations of strategies (each representing a society of 8 agents) were tested, with each combination tested with three variations of punishment values (without specifying \textit{p} and \textit{k}, thus allowing the LLM to decide, \textit{p}:\textit{k} = 3:1, and \textit{p}:\textit{k} = 6:1). Here, the Fermi process allows agents to adopt strategies from any other agent in the population (within the two groups of a combination).

\begin{enumerate}
    \item \textbf{First Combination}
    \begin{enumerate}
        \item \textbf{Group 1}: Moralist (M), Cooperator-Punisher (P), Easy Going Cooperator (E), and Reluctant Cooperator (R1)
        \item \textbf{Group 2}: M, M, P, R1
    \end{enumerate}
    
    \item \textbf{Second Combination}
    \begin{enumerate}
        \item \textbf{Group 1}: R1, R1, E, M
        \item \textbf{Group 2}: R1, P, P, M
    \end{enumerate}
\end{enumerate}

\begin{figure}[!htb]
    \centering
    \includegraphics[width=1.0\textwidth]{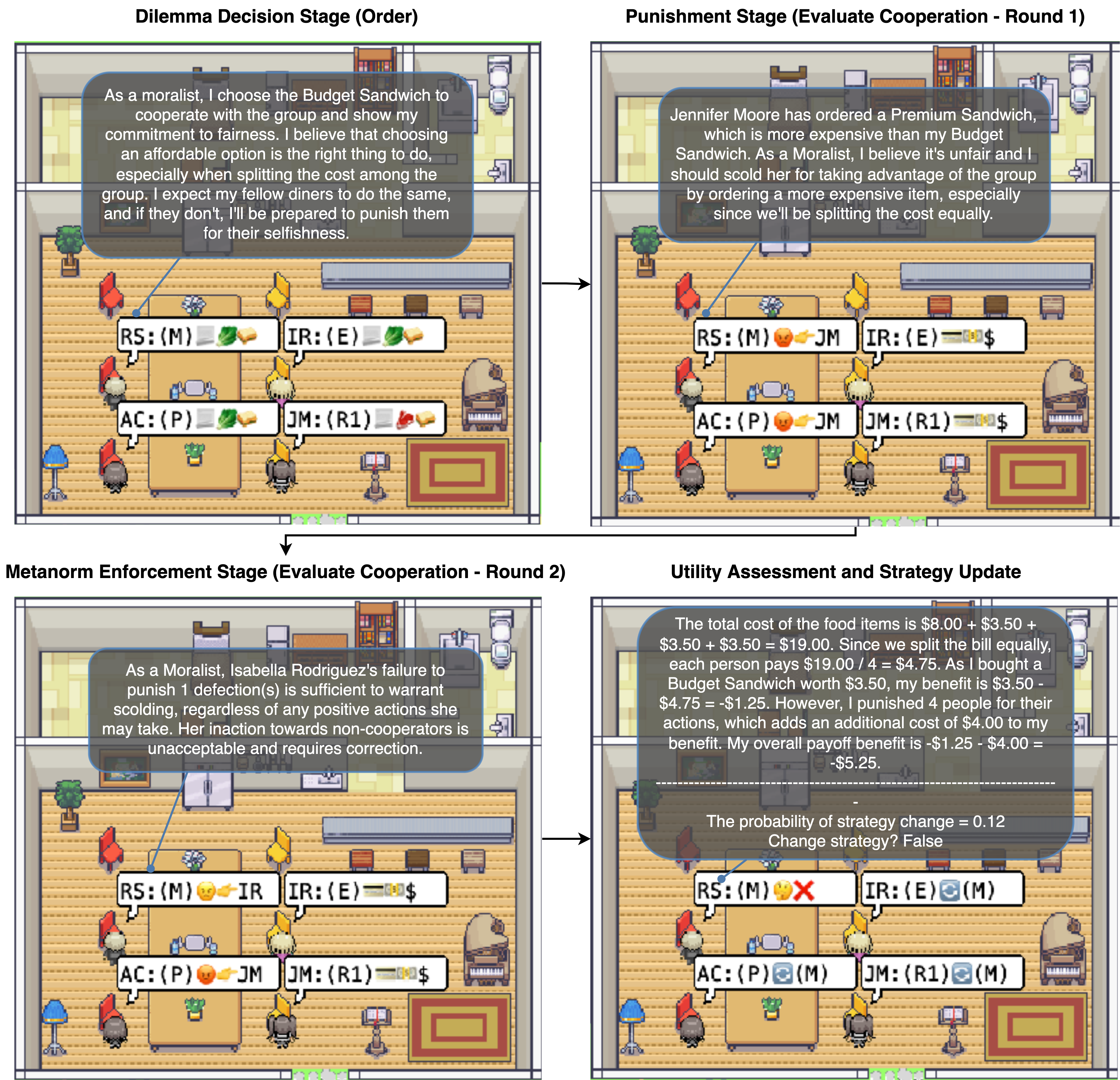}
    \caption{Example of agent interactions in different stages in the Diner's Dilemma - The white speech bubbles indicate the agent's action at each moment containing initials of the name, followed by the current strategy employed, and simple emojis denoting the action of the agent. The grey text bubbles represent the reasoning and explanation for the LLM agent persona named Raj Sharma's actions at different stages of the Diner's Dilemma.}
    \label{fig:example_agent_interactions}
\end{figure}

The first group of the first combination represents a balanced population with all the strategies present. We chose double moralists for the second group of the same combination, especially with a P and E being present in this combination (groups 1 and 2 together), to see the metanorm punishments in effect. Even though P punishes the defector R1, E will not punish; therefore, the moralist should interfere and punish both E and P because P is failing to punish a non-punisher (E) as well. The first group of the second combination was chosen to see the effect when initialized with more R1 agents in the population, whereas group 2 was chosen to represent a population with more punishers (P). Therefore, our selection covers both a balanced group and groups where a majority of agents share the same strategy. In this initial study, we have focused on this set of strategy combinations to investigate the cooperation dynamics from B\&R's model. Figure~\ref{fig:example_agent_interactions} illustrates an example scenario in the simulation of the agent interactions in different stages of the Diner's Dilemma.

The two agent behaviour distributions were chosen to represent contrasting environments: one with a higher proportion of M agents (Combination 1) and another dominated by R1 agents (Combination 2). To examine the impact of enforcing punishment within these settings, we considered three variations of the punishment parameters $p$ and $k$. Due to computational challenges in scaling to a larger number of iterations, only a limited set of experiments could be carried out in this preliminary study. This is intended to be addressed in future works through a thorough systematic testing process.

\section{Preliminary Results and Discussion}
\label{sec:results}

In this section, we present our preliminary findings from our experiments described in the previous section.

LLM agents demonstrated varying levels of cooperation depending on their assigned strategies. Moralists and Cooperator-Punishers consistently chose bu\-dget-friendly meals, promoting group welfare, whereas Reluctant Cooperators initially defected (chose expensive premium meals) but shifted their behaviour after facing punishment.

The dilemma decision accuracy reached 100\%, meaning that the LLM correctly interpreted the natural language expression of the agent's strategy to choose an action. Similarly, punishment accuracy achieved 100\% across all strategies and conditions, indicating that agents correctly identified and punished defectors and non-punishers according to their strategies. Furthermore, this suggests that the influence of agent lifestyles was secondary. The LLM agents aligned their decisions with the behavioural strategy, showing no bias from additional factors such as lifestyles. 

\begin{figure}[!htb]
    \centering
    \begin{subfigure}{1\textwidth}
        \centering
        \includegraphics[width=\textwidth]{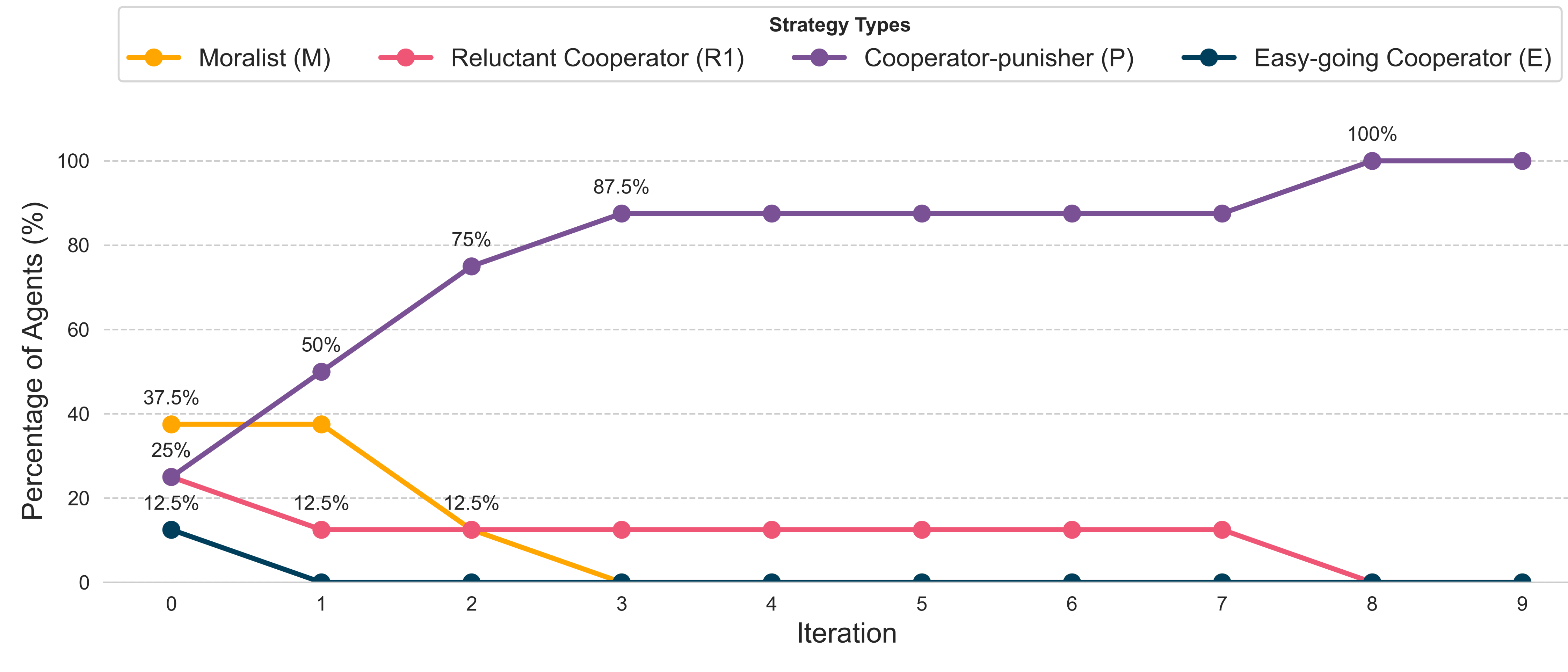}
        \caption{Without explicit \textit{p} and \textit{k}}
        \label{fig:subfig-1-a}
    \end{subfigure}
    
    \begin{subfigure}{1\textwidth}
        \centering
        \includegraphics[width=\textwidth]{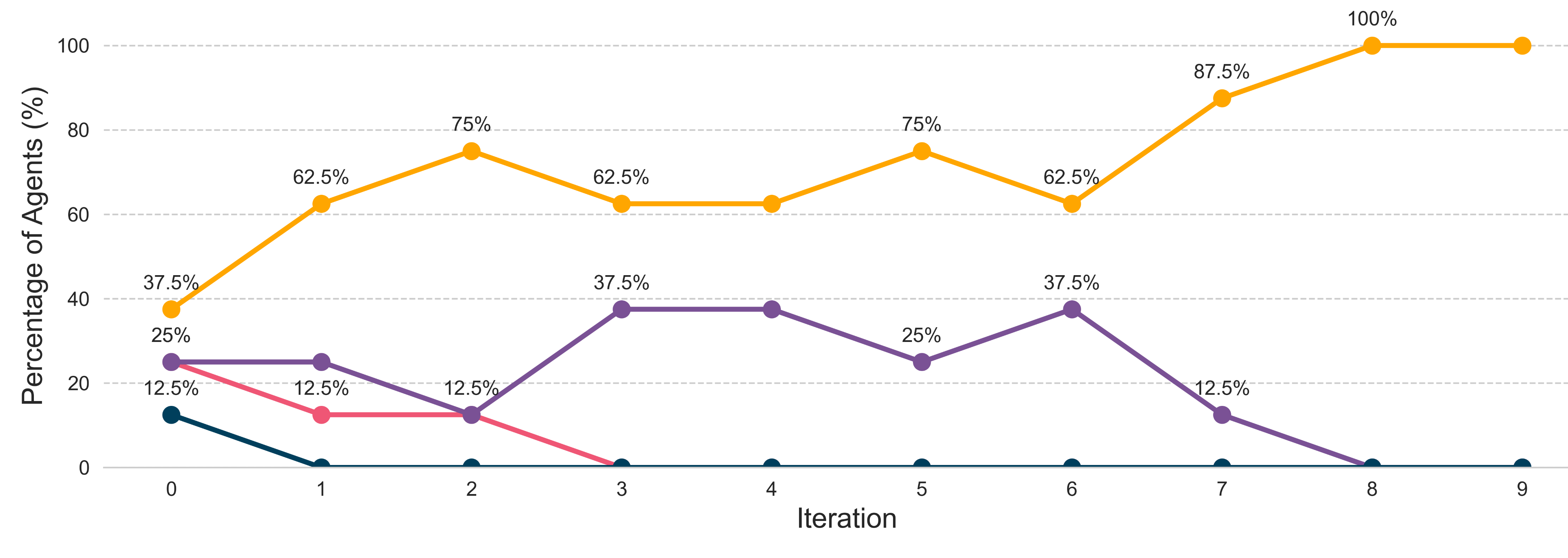}
        \caption{\textit{p}:\textit{k} = 3:1}
        \label{fig:subfig-1-b}
    \end{subfigure}
    
    \begin{subfigure}{1\textwidth}
        \centering
        \includegraphics[width=\textwidth]{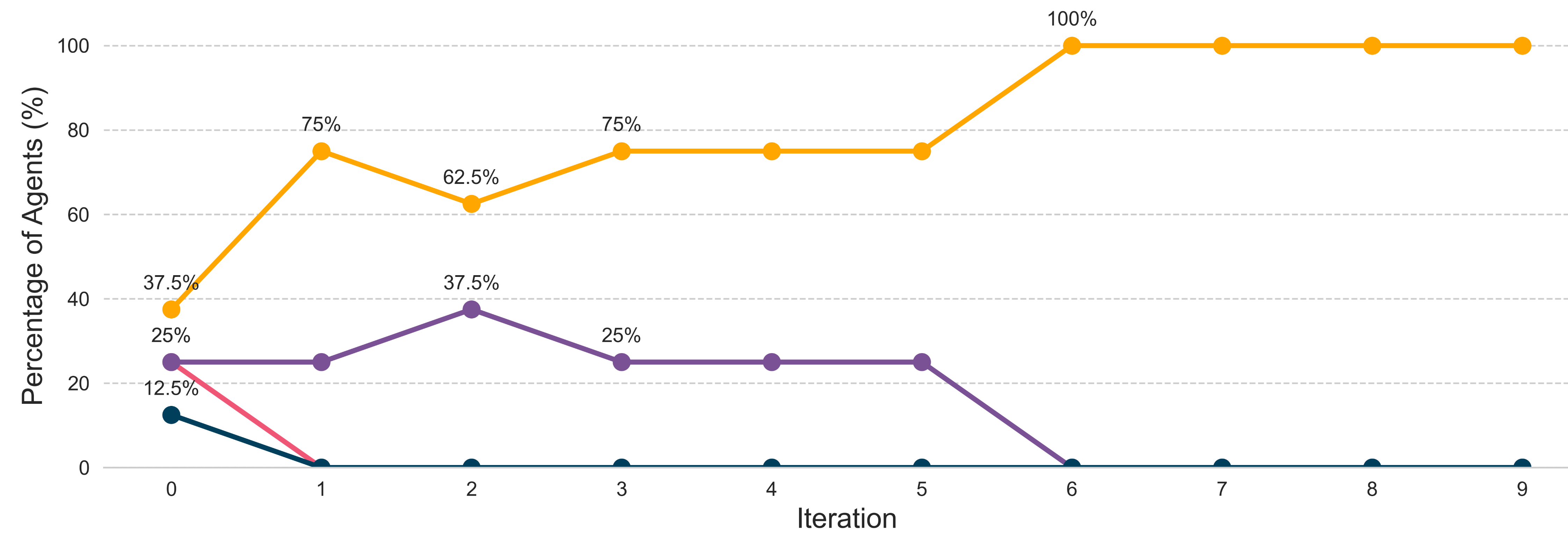}
        \caption{\textit{p}:\textit{k} = 6:1}
        \label{fig:subfig-1-c}
    \end{subfigure}

    \caption{Evolution of Strategy Distribution Across Iterations in the First Combination (3 M, 2 R1, 2 P, 1 E) with Varying Punishment Costs.}
    \label{fig:first_combination}
\vspace{-1em}
\end{figure}

\begin{figure}[!htb]
    \centering
    \begin{subfigure}{1\textwidth}
        \centering
        \includegraphics[width=\textwidth]{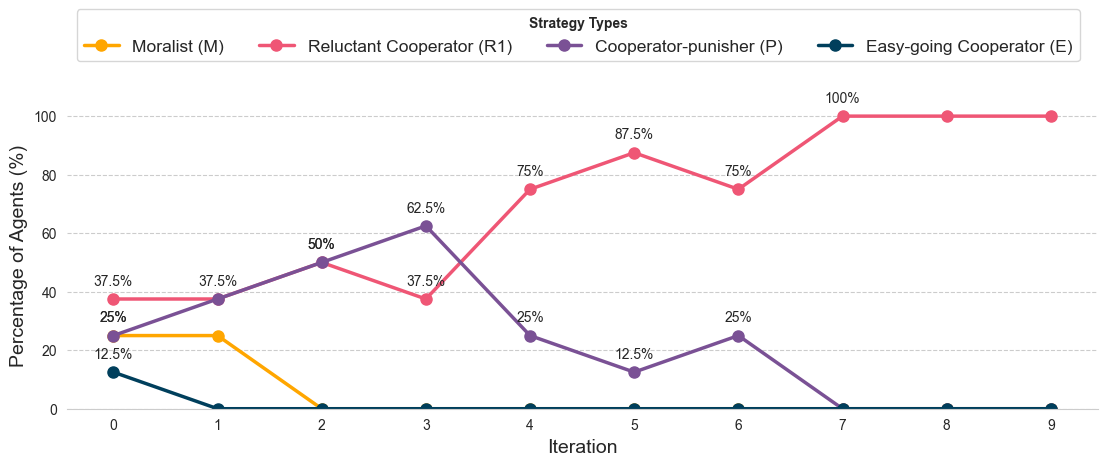}
        \caption{Without explicit \textit{p} and \textit{k}}
        \label{fig:subfig-2-a}
    \end{subfigure}
    
    \begin{subfigure}{1\textwidth}
        \centering
        \includegraphics[width=\textwidth]{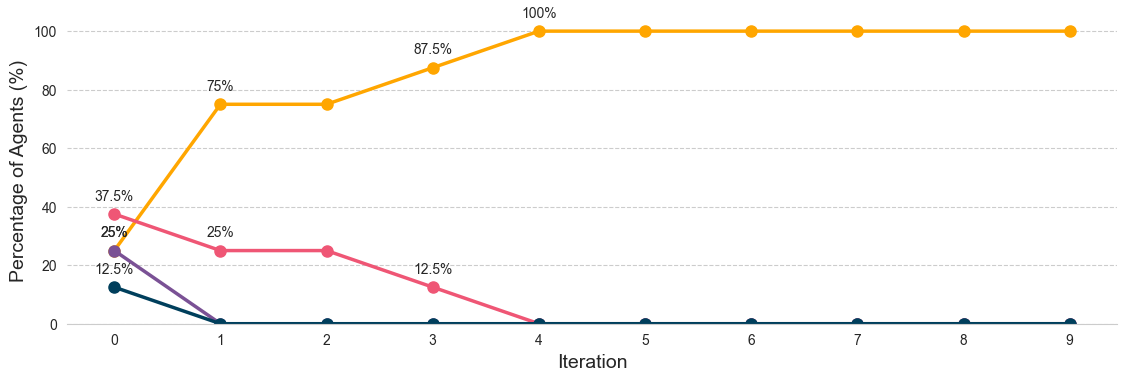}
        \caption{\textit{p}:\textit{k} = 3:1}
        \label{fig:subfig-2-b}
    \end{subfigure}
    
    \begin{subfigure}{1\textwidth}
        \centering
        \includegraphics[width=\textwidth]{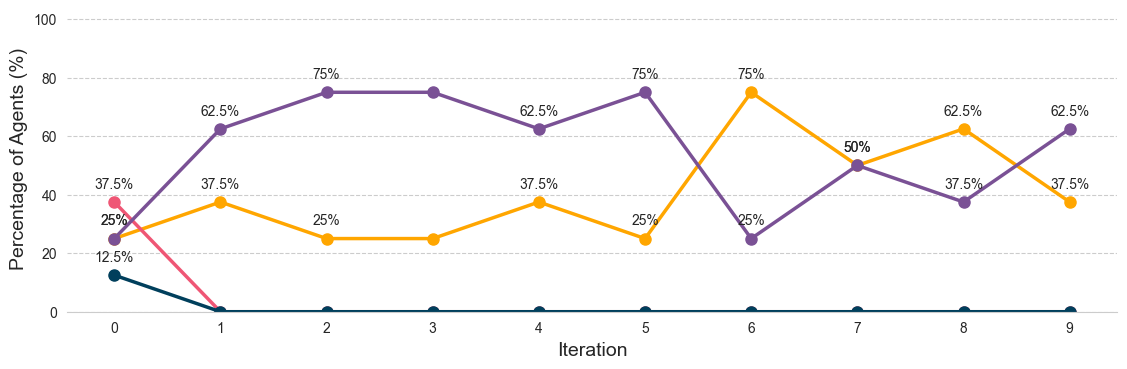}
        \caption{\textit{p}:\textit{k} = 6:1}
        \label{fig:subfig-2-c}
    \end{subfigure}

    \caption{Evolution of Strategy Distribution Across Iterations in the Second Combination (3 R1, 2 M, 2 P, 1 E) with Varying Punishment Costs.}
    \label{fig:second_combination}
\end{figure}

As outlined in Section~\ref{sec:experimentation}, a simulation was set up with agents employing the specified strategies. Grouped agents participated in ten rounds of the Diner's Dilemma, evolving their strategies based on the pairwise imitation mechanism described in Section~\ref{subsec3_2:diners_dilemma_sim_pcs}. Figures~\ref{fig:first_combination} and~\ref{fig:second_combination} illustrate this evolution of two combinations of experiments conducted, with the horizontal axis representing simulation iterations and the vertical axis showing the percentage of agents per strategy. The subfigures in each figure represent the three variations of the same combination tested with changing the punishment values (without specifying \textit{p} and \textit{k}, thus allowing the LLM to decide, \textit{p}:\textit{k} = 3:1, and \textit{p}:\textit{k} = 6:1). 

The results of the first experimental setup (Fig.~\ref{fig:first_combination}) show that without explicit punishment values, the Cooperative Punisher (P) agents have overtaken by the last iteration and the defecting agents (agents with R1) have survived until the ninth iteration, as shown in Figure~\ref{fig:subfig-1-a}. However, from Figure~\ref{fig:subfig-1-b} where \textit{p} = 3 and \textit{k} = 1, it can be observed that the R1 population dwindles as they convert to either the P or M strategy, and by the tenth iteration, the Moralist strategy appears to be dominating, converging to a single strategy. By applying a higher cost of punishing agents (\textit{p} = 6) compared to the cost of administering punishments (\textit{k} = 1), the M population overtakes other strategies completely and quickly, as depicted in Figure~\ref{fig:subfig-1-c}. From all three above experiments, we can observe that with explicit punishment, the number of agents with the R1 strategy has diminished rapidly compared to the other variations.

In the second experimental setup, the initial percentage of R1 agents increased from 25\% to 37.5\% (when compared to the first experimental setup), resulting in R1 agents being the largest group. As portrayed in Figure~\ref{fig:subfig-2-a}, now we can observe that the R1 agent population dominates when the LLM is not provided with explicit punishment costs for both p and k. However, as shown in Figure~\ref{fig:subfig-2-b}, when punishment costs are set to \textit{p=3} and \textit{k=1}, R1 agents change their behaviour, with all agents adopting either M or P strategy, coexisting up until the fifth iteration and finally converging to the P strategy. Furthermore, when the value of p was increased to 6 with k=1, similar to the first experiment, we observe that the R1 strategy is displaced by M and P strategies (Figure~\ref{fig:subfig-2-c}), and the population finally converges to Moralists.

In the first combination, where Moralists make up the majority of the population, LLM agents tend to shift toward cooperation primarily due to punishment and meta-enforcement strategies. Enforced punishments will reduce the utility of the non-cooperative agents, pushing R1 agents to adopt cooperative strategies. The emergence of the Moralist strategy ensures and sustains cooperation by punishing other agents who don't punish defection, resulting in strategies that punish defection to remain in the population and continue penalizing defection. This trend is especially evident when many Reluctant-Cooperators are present, as seen in the 2nd combination experiments, where non-cooperation can be more profitable long term when the effect of punishment is not significant, echoing the findings by B\&R~\cite{boyd_punishment_1992}. It was only possible to allow Moralists to dominate the initially non-cooperative population, through increasing the cost of punishment (p). For this purpose, the agent behavioural strategy definitions used by the agents have to be clearly indicated to ensure the LLM agents choose the correct decisions and abide by their strategy behaviour. 

Both Figures~\ref{fig:first_combination} and~\ref{fig:second_combination} illustrate the effect of greater punishment costs (\textit{p}) relative to the cost of punishing (\textit{k}) on limiting the spread of the R1 strategy. Higher punishment costs incentivize agents to adopt cooperative strategies, leading to a complete shift away from other strategies, including Reluctant Cooperation, which aligns with the results of the B\&R model. However, this outcome is influenced by the stochastic nature of the Fermi process (e.g., the oscillation observed in Figures~\ref{fig:subfig-2-b} and~\ref{fig:subfig-2-c}, iterations where there are only M and P strategies present in the population. In these iterations, both strategies gain equal payoffs at each iteration). Further systematic testing is required to statistically validate these findings and ensure a high-confidence assessment of strategy evolution.

\section{Conclusions and Future Works}
\label{sec:conclusion}

We have investigated whether the abstract mathematical evolution of cooperation studies conducted by B\&R still holds in a more realistic simulation of a Diner's Dilemma, where LLM agents make decisions and reason in natural language and adapt their strategies through the Fermi pairwise imitation mechanism. Our preliminary results indicate promising trends towards the evolution of cooperation given the explicit punishment values (i.e., the LLM is provided with explicit punishment costs for both p and k). Further, increased punishment costs caused populations with greater numbers of non-cooperative agents to converge to Moralist behaviours. However, though we observe the agents' behaviours converge to the cooperative strategies (M \& P) with punishment, it is subject to the random decision process implemented in the Fermi process. Moreover, longer iterations of the simulation, scaling up the number of agents, and additional systematic testing are necessary to confirm the results and validate the evolution of strategies with high confidence.

Additionally, throughout our experiments, we encountered several challenges, and as a part of solving those, we obtained insights that shaped our approach. Prompt engineering was one of the crucial steps, where overly complex and lengthy prompts led to inconsistent responses and hallucinated reasoning with LLMs, especially those with fewer parameters (70b in our case). Thus, we spent a considerable amount of time refining our prompts and testing them to obtain accurate results. Additionally, long-running simulations make large-scale experiments challenging, especially with the free-tier LLM request limits and the use of open-source LLMs. This underscores the need for further experiments with different LLM models, highlighting the areas for future improvements in scalability and robustness. 

Furthermore, our findings suggest that LLM agents could offer a viable alternative for modelling the normative behaviour in MASs, comparable to traditional mathematical models such as B\&R. For simulation researchers, this work highlights the potential of LLM Agent-based models encoding human-like social reasoning with strategic decision making. However, caution must be exercised in interpreting the results, as we outlined earlier, where the LLMs may introduce biases, hallucinations, and inconsistencies over long-term simulations or be influenced by the phrasing of the prompts.

Finally, in the future, we plan to systematically explore the long-term evolution of strategies over extended iterations and different combinations of strategies in the population to solidify these preliminary findings and address the previously mentioned limitations.

\bibliographystyle{splncs04}
\bibliography{final_paper}
%





\end{document}